# Improving Quality of Clustering using Cellular Automata for Information retrieval

[1]P. Kiran Sree, [2]G.V.S. Raju, [3]I. Ramesh Babu and [4]S. Viswanadha Raju
[1]Department of CSE, S.R.K.Institte of Technolgy, Vijayawada, India
[2]Department of CSE, Swarnandhara Engineering College, Narasapur, India
[3]Department of CSE, Acharya Nagarjuna University, Guntur, India
[4]Department of CSE, Gokaraju Rangaraju Institute of Engg and Technology, India

**Abstract:** Clustering has been widely applied to Information Retrieval (IR) on the grounds of its potential improved effectiveness over inverted file search. Clustering is a mostly unsupervised procedure and the majority of the clustering algorithms depend on certain assumptions in order to define the subgroups present in a data set .A clustering quality measure is a function that, given a data set and its partition into clusters, returns a non-negative real number representing the quality of that clustering. Moreover, they may behave in a different way depending on the features of the data set and their input parameters values. Therefore, in most applications the resulting clustering scheme requires some sort of evaluation as regards its validity. The quality of clustering can be enhanced by using a Cellular Automata Classifier for information retrieval. In this study we take the view that if cellular automata with clustering is applied to search results (query-specific clustering), then it has the potential to increase the retrieval effectiveness compared both to that of static clustering and of conventional inverted file search. We conducted a number of experiments using ten document collections and eight hierarchic clustering methods. Our results show that the effectiveness of query-specific clustering with cellular automata is indeed higher and suggest that there is scope for its application to IR.

**Key words:** Cellular automata, information retrieval, clustering

## INTRODUCTION

Locating interesting information is one of the most important tasks in Information Retrieval (IR). An IR system accepts a query from a user and responds with a set of documents. The system returns both relevant and non-relevant material and a document organization approach is applied to assist the user in finding the relevant information in the retrieved set. Generally a search engine presents the retrieved document set as a ranked list of document titles. The documents in the list are ordered by the probability of being relevant to the user's request. The highest ranked document is considered to be the most likely relevant document; the next one is slightly less likely and so on. This organizational approach can be found in almost any existing search engine. A number of alternative document organization approaches have been developed over the recent years. These approaches are normally based on visualization and presentation of some relationships among the documents, terms, or the user's query. One of such approaches is document clustering. Document clustering has been studied in the information retrieval for several decades.

Clustering is a mostly unsupervised procedure and the majority of the clustering algorithms depend on certain assumptions in order to define the subgroups present in a data set. Moreover, they may behave in a different way depending on the features of the data set and their input parameters values. Therefore, in most applications the resulting clustering scheme requires some sort of evaluation as regards its validity. Our experimental results confirm the reliability of our index showing that it performs favorably in all cases selecting independently of clustering algorithm the scheme that best fits the data under consideration

## CELLULAR AUTOMATA (CA) AND FUZZY CELLULAR AUTOMATA (FCA)

A CA[4-6], consists of a number of cells organized in the form of a lattice. It evolves in discrete space and time. The next state of a cell depends on its own state and the states of its neighboring cells. In a 3-neighborhood dependency, the next state $q_i(t+1)$ of a cell is assumed to be dependent only on itself and on its two neighbors (left and right) and is denoted as:

$$q_i(t + 1) = f(q_{i-1}(t), q_i(t), q_{i+1}(t)) \quad (1)$$

where, $q_i(t)$ represents the state of the $i^{th}$ cell at $t^{th}$ instant of time, f is the next state function and referred to as the rule of the automata. The decimal equivalent of the next state function, as introduced by Wolfram, is the rule number of the CA cell[8]. In a 2-state 3-neighborhood CA, there are total 256 distinct next state functions.

**FCA fundamentals:** FCA[2,6] is a linear array of cells which evolves in time. Each cell of the array assumes a state $q_i$, a rational value in the interval [0,1] (fuzzy states) and changes its state according to a local evolution function on its own state and the states of its two neighbors. The degree to which a cell is in fuzzy states 1 and 0 can be calculated with the membership functions. This gives more accuracy in finding the coding regions. In a FCA, the conventional Boolean functions are AND, OR, NOT.

**Dependency matrix for FCA:** Rule defined in Eq. 1 should be represented as a local transition function of FCA cell. That rules (Table 1) are converted into matrix form for easier representation of chromosomes.

**Example 1:** A 4-cell null boundary hybrid FCA with the following rule <238, 254, 238, 252> (that is, <($q_i+q_{i+1}$), ($q_{i-1}+q_i+q_{i+1}$), ($q_i+q_{i+1}$), ($q_{i-1}+q_i$)>) applied from left to right, may be characterized by the following dependency matrix.

While moving from one state to other, the dependency matrix indicates on which neighboring cells the state should depend. So cell 254 depends on its state, left neighbor and right neighbor Fig. 1. Now we represented the transition function in the form of matrix. In the case of complement[5,6,8], FMACA we use another vector for representation of chromosome.

**Transition from one state to other:** Once we formulated the transition function, we can move form one state to other. For the example 1 if initial state is P (0) = (0.80, 0.20, 0.20, 0.00) then the next states will be:

$$P(1) = (1.00\ 1.00,\ 0.20,\ 0.20)$$
$$P(2) = (1.00\ 1.00,\ 0.40,\ 0.40)$$
$$P(3) = (1.00\ 1.00,\ 0.80,\ 0.80)$$
$$P(4) = (1.00\ 1.00,\ 1.00,\ 1.00)$$

**Search strategy:** In this research, we select the steepest descent strategy. It can make an evaluation for every solution in a neighborhood of P, then choose one which can make the objective criterion function have maximal gain as a new solution. That is to say, it searches a candidate solution that can improve results furthest.

Suppose neighborhood of P is Neighbour (P). The steepest descent strategy is to search a P` P` = Argmax (E(P`)-E(P)|P`∈ Neighbour(P)) in Neighbour(P).

For any p∈ Neighbour(P), E(P`)≥E(P) and E(P`)-E(P)>0.

Table 1: FA rules

| Non-complemented rules | | Complemented rules | |
|---|---|---|---|
| Rule | Next state | Rule | Next state |
| 0 | 0 | 255 | 1 |
| 170 | $q_i+1$ | 85 | $\overline{q_i+1}$ |
| 204 | $q_i$ | 51 | $\overline{q_i}$ |
| 238 | $q_i+q_{i+1}$ | 17 | $\overline{q_i+q_{i+1}}$ |
| 240 | $q_{i-1}$ | 15 | $\overline{q_{i-1}}$ |
| 250 | $q_{i-1}+q_{i+1}$ | 5 | $\overline{q_{i-1}+q_{i+1}}$ |
| 252 | $q_{i-1}+q_i$ | 3 | $\overline{q_{i-1}+q_i}$ |
| 254 | $q_{i-1}+q_i+q_{i+1}$ | 1 | $\overline{q_{i-1}+q_i+q_{i+1}}$ |

$$T = \begin{bmatrix} 1 & 1 & 0 & 0 \\ 1 & 1 & 1 & 0 \\ 0 & 0 & 1 & 1 \\ 0 & 0 & 1 & 1 \end{bmatrix}$$

Fig. 1: Matrix representation

$$E(S_i - \{d\}) - E(S_i) = \|D_i - d\| - \|D_i\|$$
$$= \left(\sqrt{\|D_i\|^2 - 2\|D_i\|d \cdot c_i + 1} - \|D_i\|\right) \quad (2)$$

$$E(S_j \cup \{d\}) - E(S_j) = \|D_j + d\| - \|D_j\|$$
$$= \left(\sqrt{\|D_j\|^2 + 2\|D_j\|d \cdot c_j + 1} - \|D_j\|\right) \quad (3)$$

The text clustering algorithm based on Cellular automata based local search (TCLS) algorithm is composed of the following steps (one step):

- For one clustering partition P = $(S_1, S_2,...,S_K)$
- Suppose max Δ = 0, movedDoc = null, target = null, (movedDoc is the text that need to be moved, target is the target class that movedDoc moves to);
  For every text d∈ $S_i$ in S
  For all j, j = 1,2…,K ∧ j ≠ i
  calculate $\Delta_j E(P) \equiv E(P') - E(P)$

  In P, let text d moves from $S_i$ to $S_j$, then get the P`
  Let b = argmax {$\Delta_j$ E (P) > 0| j ≠ i}

$$\max\Delta = \max(\max\Delta, b), \text{movedDoc} = d, \text{target} = S_j$$

- If movedDoc ≠ null (a best optimal solution has already been found), then let d move from $S_i$ to target and recalculate $D_i$ and $D_{target}$
- Return the partition P`

The difference between Cellular automata based local search strategy and K-Means is:

$$\text{Suppose } P = (S_1, S_2, \ldots, S_K)$$

P` = (S`$_1$, S`$_2$, …, S`$_K$) ∈ Neighbour (P) the target is.

**Clustering algorithm based on cellular automata based local search:** Cellular automata based local search Based Clustering (LSC):

- Give an initial clustering partition $P = (S_1, S_2, \ldots, S_K)$
- Run TCLS
- If satisfy stop condition, then exit, else run step 2

When the algorithm is running, the required space and time in every iterative are the same, the bottleneck of calculation is to calculate $\Delta_{P'}E(P) \equiv E(P') - E(P)$, that is for any $d \in S_i$:

$$\text{Calculate: } E(S_i - \{d\}) - E(S_i)$$

and

$$E(S_j \cup \{d\}) - E(S_j)$$

It can be noted that, in every iterative of K-Means, the $\|D_i\|$ and $d-c_i$, $d \in S$, $i = 1, 2, \ldots, K$ are all need to be calculated.

That is to say, the time, space and calculate complexity in every iterative of Cellular automata based local search or K-Means are almost the same.

$$\Delta_{P'}E(P) = \sum_{d \in S_i'} d - (c_i' - c_i') + \sum_{d \in S_i'} d - (c_i' - c_i') + \Delta_K \quad (4)$$

$$\Delta_{P'}E(P) \geq \Delta_k \quad (5)$$

**Information retrieval system evaluation:** To measure ad hoc IR effectiveness in the standard way, we need a test collection consisting of three things:

- A document collection
- A test suite of information needs, expressible as queries
- A set of relevance judgments, assessment of either relevant or non relevant for each query-document pair

**EXPERIMENTAL RESULTS**

In order to evaluate the classification without taking into account the position of clusters, we use the list of relevant documents supplied by NIST for TREC-7 and, for each query, select the best clusters according to the number of relevant documents they contain. Hence, we can measure how much the classification groups relevant documents. Let Lq be the list of documents constructed from the succession of the clusters ranked according to the number of relevant items they contain. The evaluations presented below have been obtained by means of the trec_eval application over the 50 queries (351-400) of TREC-7. The corpus was the one used for TREC-7 (528.155 documents extracted from the Financial Times, Federal Register, Foreign Broadcast Information Service and the LA Times. We have used the IR system developed by LIA and Bertin and Cie to obtain the lists of documents to cluster. Whenever possible, the first 1000 documents retrieved for each query have been kept to cluster them.

**Number of classes:** The number of classes is defined at the start of the process. It cannot grow but can be reduced when a class empties. In next figures, the indicated numbers of clusters correspond to the values initially chosen. The documents that are not assigned to a class at the end of the classification are allocated to a new one (at the last position in the ranked list of clusters). By choosing the same number of classes from 2-13 for all queries, the levels of the average precision over all relevant documents are lower than those without classification with lists.

The differences between results indicated in Fig. 2 and 3 measure how much the above defined distance 4 We choose ni > ni+1 to favor the first ranked classes ranks the clusters. The average precision decrease is about 5% when clusters are ranked according to the computed distances and not according to the number of relevant documents they contain.

Let $L_q$ be the list of documents constructed from the succession of the clusters ranked according to the number of relevant items they contain.

Let $L_c$ be the list of documents constructed from the succession of the clusters ranked according to their distances with the query using CA classifier:

$$L_c = C1 \times C2 \times C3 \times K$$

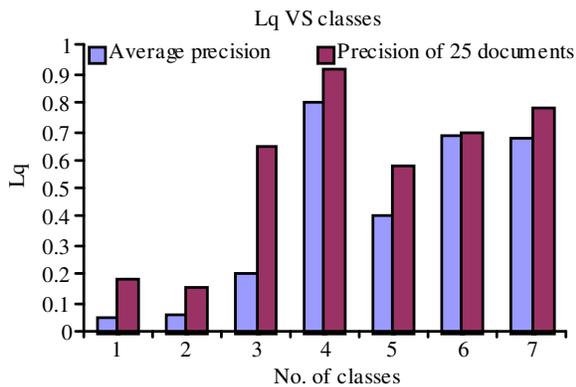

Fig. 2: $L_q$ with different numbers of classes (precision at 10 and average precision without classification are respectively indicated)

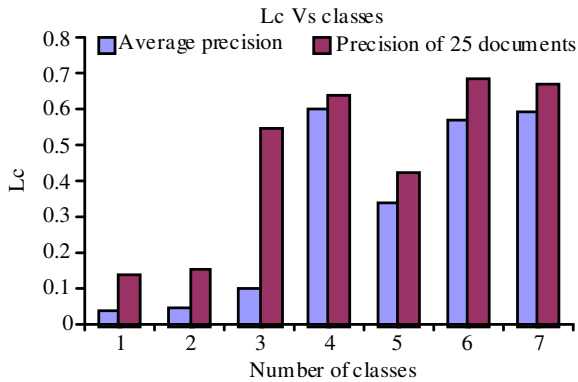

Fig. 3: $L_C$ with different numbers of classes

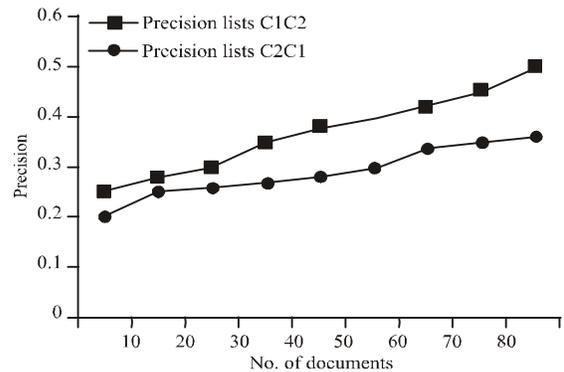

Fig. 4: Precision of lists C1C2 and C2C1 (Pure Clustering)

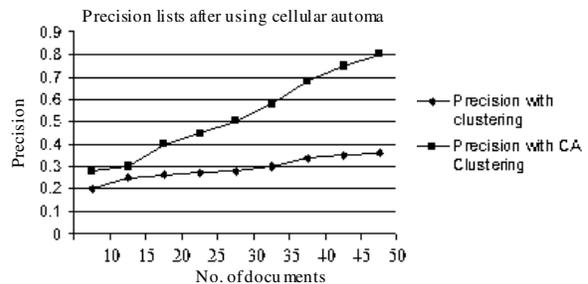

Fig. 5: Precision using CA classification

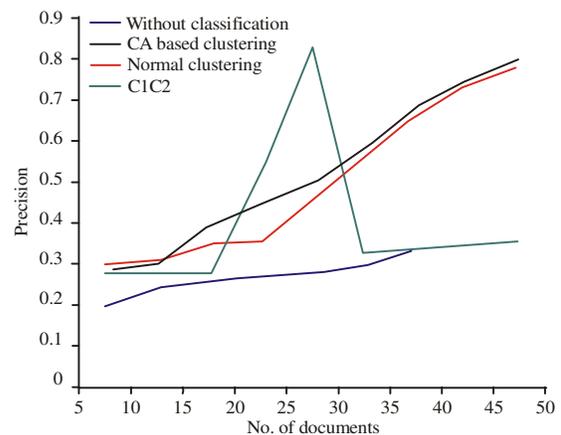

Fig. 6: Precision for different browsing aspects

By choosing the same number of classes from 5-25 for all queries, the levels of the average precision over all relevant documents are lower than those without classification with lists $L_q$ and $L_C$ (Fig. 2 and 3). The decrease rate varies from 2.2-3%. Figure 2 and 3 show that those lists do not allow to globally improve results of the retrieval. The average precision decreases since the relevant documents that are not in the first cluster are ranked after all items of that one. The differences between results indicated in Fig. 2 and 3 measure how much the above defined distance 4 We choose ni > ni+1 to favor the first ranked classes. ranks the clusters. The average precision decrease is about 5% when clusters are ranked according to the computed distances and not according to the number of relevant documents they contain.

However, the first ranked cluster according to the distances to the queries is very often better than the next ones as shown in Fig. 4 where we have compared lists C1C2 and C2C1. With this second list, the relative decrease of the average precision over the 145 queries equals 28% (from 0.21 to 0.089). In Fig. 5, we can see that the first ranked cluster is the best 3 times for the 5 queries indicated among 6 clusters.

Figure 5 and 6 show some results obtained with 203 clusters. One can see that precision at low level of recall with lists Ln are better than those of list LC (succession of each cluster's contents). However, only list Lq allows to obtain better results than without classification. At recall 0.21, the relative increase of precision of list L5 over list LC equals 19.5% (from 0.297 to 0.3892). Figure 5 and 6 shows the results obtained by using the title field as queries and then the whole topic (list LCn with 2 clusters). Not surprisingly, the best results are obtained with longer queries even if in some cases the narrative field contains words not wanted.

## CONCLUSION

We see that clustering with CA with local search can greatly improve the effectiveness of the ranked list. In fact it can be as effective as the interactive relevance feedback based on query expansion. Surprisingly this high performance can be achieved by following a very simple strategy. Given a list of clusters created by the CA based local search algorithm starts at the top of the list and follows it down examining the documents in each cluster. The experimental results proves the improvement of clustering quality with addition o f Cellular Automata.

It increases the effectiveness of retrieval by providing to users at least one cluster with a precision higher than the one obtained without using CA. We have examined, with TREC-7 corpora and queries, the impact on the classification results of the cluster numbers and of the way to browse them. We have shown that a variation of the number of clusters according to the query size improves the results. By automatically constructing a new ranked list according to the distances between clusters and queries, the precision is lower than without CA. The evaluation of other distances is in progress.